\documentclass[journal]{IEEEtran}

\ifCLASSINFOpdf
  \usepackage[pdftex]{graphicx}
	\graphicspath{ {images/} }

\else

\fi

\usepackage{amsmath}

\usepackage{url}
\usepackage{cite}
\usepackage{blindtext}
\usepackage{caption}
\usepackage{subcaption}
\usepackage[dvipsnames]{xcolor}
\usepackage{amssymb}
\usepackage{enumitem}
\usepackage{lipsum}
\usepackage{mathtools}
\usepackage{mathtools, cuted}
\usepackage{lipsum, color}
\usepackage[margin=15mm]{geometry}
\usepackage{multicol}
\usepackage{float}
\usepackage{subcaption}
\usepackage[ruled,vlined]{algorithm2e}
\usepackage{algorithmic}
\usepackage{etoolbox}
\patchcmd{\thebibliography}
  {\settowidth}
  {\setlength{\itemsep}{0pt plus 0.1pt}\settowidth}
  {}{}
\apptocmd{\thebibliography}
  {\small}
  {}{}

\floatstyle{plain}
\newfloat{twocolequfloat}{b}{zzz}
\floatname{twocolequfloat}{Equation}

\begin{document}

\title{An Adaptive Receiver for Underwater Acoustic Full-Duplex Communication with Joint Tracking of the Remote and Self-Interference Channels\\

\thanks{\noindent\rule{8.8cm}{0.4pt}

This work was supported primarily by the Engineering Research Centers Program of the National Science Foundation under NSF Cooperative Agreement No. (CNS-1704097, CNS-1704076). Any opinions, findings and conclusions or recommendations expressed in this material are those of the authors and do not necessarily reflect those of the National Science Foundation.}
}

\author{\IEEEauthorblockN{Mohammad Towliat$^1$,
Zheng Guo$^2$,
Leonard J. Cimini$^3$,
Xiang-Gen Xia$^4$, and
Aijun Song$^5$
} \vspace{10pt}\\

\IEEEauthorblockA{$^{1, 3, 4} $\textit{Department of Electrical and Computer Engineering }\\
\textit{University of Delaware}\\
Newark, DE, USA \\
\{mtowliat, cimini, xianggen\}@udel.edu \vspace{10pt}}\\
\IEEEauthorblockA{$^{2, 5} $\textit{Department of Electrical and Computer Engineering} \\
\textit{University of Alabama}\\
Tuscaloosa, AL, USA \\
zguo18@crimson.ua.edu, song@eng.ua.edu}
\vspace{-20pt}
}

%
%

\markboth{}%
{Shell \MakeLowercase{\textit{et al.}}: Bare Demo of IEEEtran.cls for IEEE Journals}
%



\maketitle

\begin{abstract}
Full-duplex (FD) communication is a promising candidate to address the data rate limitations in underwater acoustic (UWA) channels. Because of transmission at the same time and on the same frequency band, the signal from the local transmitter creates self-interference (SI) that contaminates the the signal from the remote transmitter. At the local receiver, channel state information for both the SI and remote channels is required to remove the SI and equalize the SI-free signal, respectively.  However, because of the rapid time-variations of the UWA environment, real-time tracking of the channels is necessary. In this paper, we propose a receiver for UWA-FD communication in which the variations of the SI and remote channels are jointly tracked by using a recursive least squares (RLS) algorithm fed by feedback from the previously detected data symbols. Because of the joint channel estimation, SI cancellation is more successful compared to UWA-FD receivers with separate channel estimators. In addition, due to providing a real-time channel tracking without the need for frequent training sequences, the bandwidth efficiency is preserved in the proposed receiver. 
\end{abstract}


%
\IEEEpeerreviewmaketitle

\vspace{-5pt}
\section{Introduction}
%
%
%
%
\IEEEPARstart{L}{imited} available bandwidth and dynamic transmission media are two serious issues that challenge underwater acoustic (UWA) communication systems and lead to a data transmission rate which does not exceed a few tens of kilobits per second \cite{R1}.  Full-duplex (FD) communication is an interesting technique to increase the data rate in UWA channels. By performing transmission and reception at the same time and on the same frequency band, UWA-FD has the potential to double the data rate of half-duplex systems. However, it is challenging to eliminate the self-interference (SI) caused by the local transmitter. A combination of analog and digital  methods can be used for SI cancellation \cite{R2}; it has also been shown that fully-digital SI cancellation methods are capable of minimizing SI \cite{R3}.
In fully-digital approaches, assuming knowledge of the SI channel, the receiver is able to estimate and eliminate the SI; then, the SI-free signal can be equalized with the purpose of detecting the data symbols from a remote transmitter. Obviously, this equalizer is designed to correct for the impulse response of the remote channel that links the receiver and the remote transmitter. So, channel state information for both the SI and remote channels is required at the receiver; however, since the SI signal is much more powerful than the remote signal, an accurate tracking of the SI channel is critical.
 
Because of fluctuations in the UWA transmission media, the SI and remote channels are time-varying \cite{R10, R9}, and estimates are valid  for just a short time, roughly equivalent to the coherence time of the channel. So, real-time tracking of both time-varying channels is necessary. Addressing this concern, in this paper, we propose an adaptive UWA-FD receiver with three stages of processing at each cycle: channel estimation, SI cancellation, and equalization. At the channel estimation stage, the SI and remote channels are jointly estimated; at the next two stages, these estimates are used to cancel the SI and equalize the SI-free signal, respectively. 

There are a few recent related works using the adaptive filtering approach to perform SI cancellation in UWA channels \cite{R4, R11}. In these works, only the SI channel is estimated by the adaptive filters, and the remote channel is assumed to be known or estimated in a further separate step. Since the SI is much stronger than the remote signal, the SI channel is estimated by treating the remote signal as an additional additive noise. Then, the SI signal is removed from the received signal based on the estimated SI channel. Obviously, during this procedure, the SI channel estimation and, consequently, the SI removal is affected by the unknown remote signal, which leads to a performance degradation \cite{R13}. By contrast, in our proposed receiver, the SI and remote channels are jointly estimated which implies that, during the channel estimation stage, both the SI and remote signals are taken into consideration. 
As a result, one of the advantages of this approach is that the SI channel estimation is more accurate which, consequently, results in a better SI cancellation.
The other advantage of the proposed receiver is that the remote channel is tracked along with the SI channel without using any extra training sequences. This means that variations in the remote channel are tracked by the receiver without wasting the bandwidth needed for recurrent training sequences from the remote transmitter.  

The rest of this paper is organized as follows. In Section II, the SI in UWA-FD system is described and formulated. The proposed UWA-FD receiver is presented in Section III. Simulation results are given in Section IV and we conclude the paper in Section V.

\section{Problem Formulation}
\begin{figure}[t!]
\centering
\includegraphics [width=2.5in]{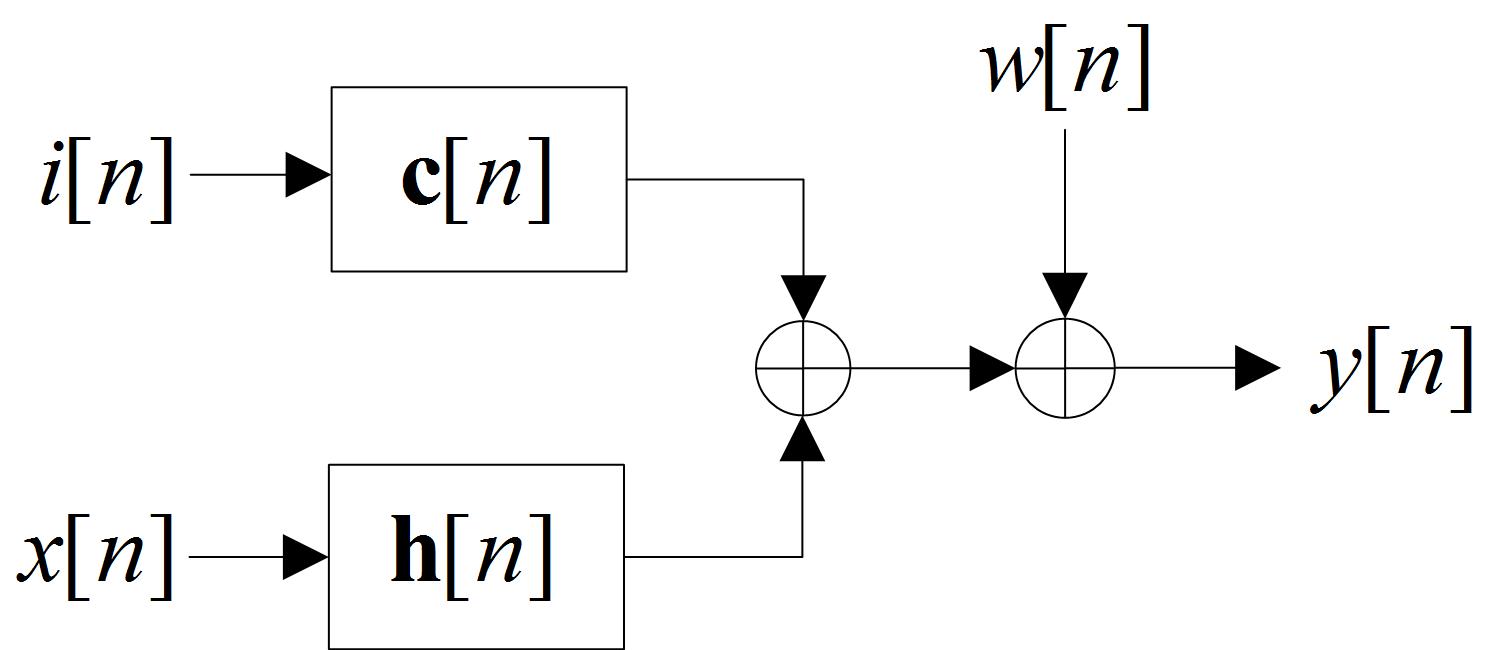}
\caption{Model for baseband interference at the receiver.}
\vspace{-10pt}
\label{F1}
\end{figure}

In Fig.~\ref{F1}, a model of the baseband interference at the reviver is shown; the SI signal interferes with the desired remote signal. In this model, $i[n]$ and $x[n]$ are the reference sequences for the local and remote transmitters, respectively. The local transmitter's reference, $i[n]$, is the downconverted, matched filtered, and downsampled (to the symbol rate) output of the local power amplifier (PA). Because of the non-linearity and noise of the PA \cite{R5}, $i[n]$ is not necessarily equal to the local transmitted symbols.
However, the remote transmitter's reference, $x[n]$, represents the transmitted symbols from the remote transmitter. The first reason for these considerations is that, since  the SI signal is relatively powerful, the non-linearity and noise effects of the local PA are not negligible at the receiver; by contrast, because  of the lower power of the remote signal at the receiver, the distortions of the remote PA are considered trivial \cite{R6}. Secondly, it must be noted that the pulse shapes that are used to shape the transmitted symbols at the local transmitter are not aligned with the matched filter at the receiver because the receiver's matched filter is paired with the pulse shapes from the remote transmitter. The misalignment between the receiver and the local transmitter causes an additional distortion on the transmitted symbols from the local transmitter. In order to take this into account, the  matched filter that is used to generate $i[n]$ must be aligned with the receiver, not with the local transmitter. According to the above discussion, $i[n]$ can take any soft value, whereas, $x[n]$ is the actual data symbol generated by the remote modulator. 

In the model presented in Fig.~\ref{F1}, in order to consider the multipath fading in the UWA transmission media, $\mathbf{c}[n]={{\left[ {{c}_{0}}[n],\ldots {{c}_{M-1}}[n] \right]}^{H}}$and $\mathbf{h}[n]={{\left[ {{h}_{\text{0}}}[n],\ldots {{h}_{L-1}}[n] \right]}^{H}}$ are the  time-varying SI and remote channel vectors with length $M$ and $L$, respectively. By defining the signal vectors  $\mathbf{i}[n]={{\left[ i[n],\ldots i[n-M+1] \right]}^{T}}$and $\mathbf{x}[n]={{\left[ x[n],\ldots x[n-L+1] \right]}^{T}}$, the received signal can be represented as
\begin{equation}
y[n]=s[n]+r[n]+w[n], \label{1}
\end{equation}
where $w[n]$ denotes the additive ambient noise with power $\sigma _0^2$; $s[n]$ and $r[n]$ are respectively the SI and remote signals, with power $P_s$ and $P_r$, and can be defined as
\begin{align}
 s[n]&={\mathbf{c}}^{H}[n]\,\mathbf{i}[n], \nonumber\\ 
 r[n]&={\mathbf{h}}^{H}[n]\,\mathbf{x}[n].
\label{2} 
\end{align}
Obviously, in \eqref{1} and \eqref{2}, it is assumed that the channel coherence time is long enough that both the SI and remote channels remain invariant over $K=\max \{ M,L\}$ time instances.  

The ultimate goal at the receiver is to detect the data symbol $x[n]$ from the received signal. Since $i[n]$ is known, the simplest approach is to determine $s[n]$ and remove it from the received signal, $y[n]$, and then employ an equalizer to detect the data symbols $x[n]$ from the SI-free signal. 
An estimate of the SI channel is required for SI cancellation, and an estimate of the remote channel is needed for equalization. On the other hand, because of fluctuations in the UWA transmission media, both the SI and remote channels are time-varying and, hence, the receiver should be able to keep track of these changes without wasting the bandwidth caused by using frequent training intervals. 
To respond to this concern, in this paper, we propose a receiver in which the variations of the SI and remote channels are jointly tracked without requiring recurrent training sequences from the remote transmitter. We remove the necessity of transmitting training by taking feedback from the previously detected data symbols and using it as a reference sequence for the remote transmitter.

\vspace{-5pt}
\section{The proposed UWA-FD receiver}

\begin{figure}[t!]
\centering
\includegraphics [width=2in]{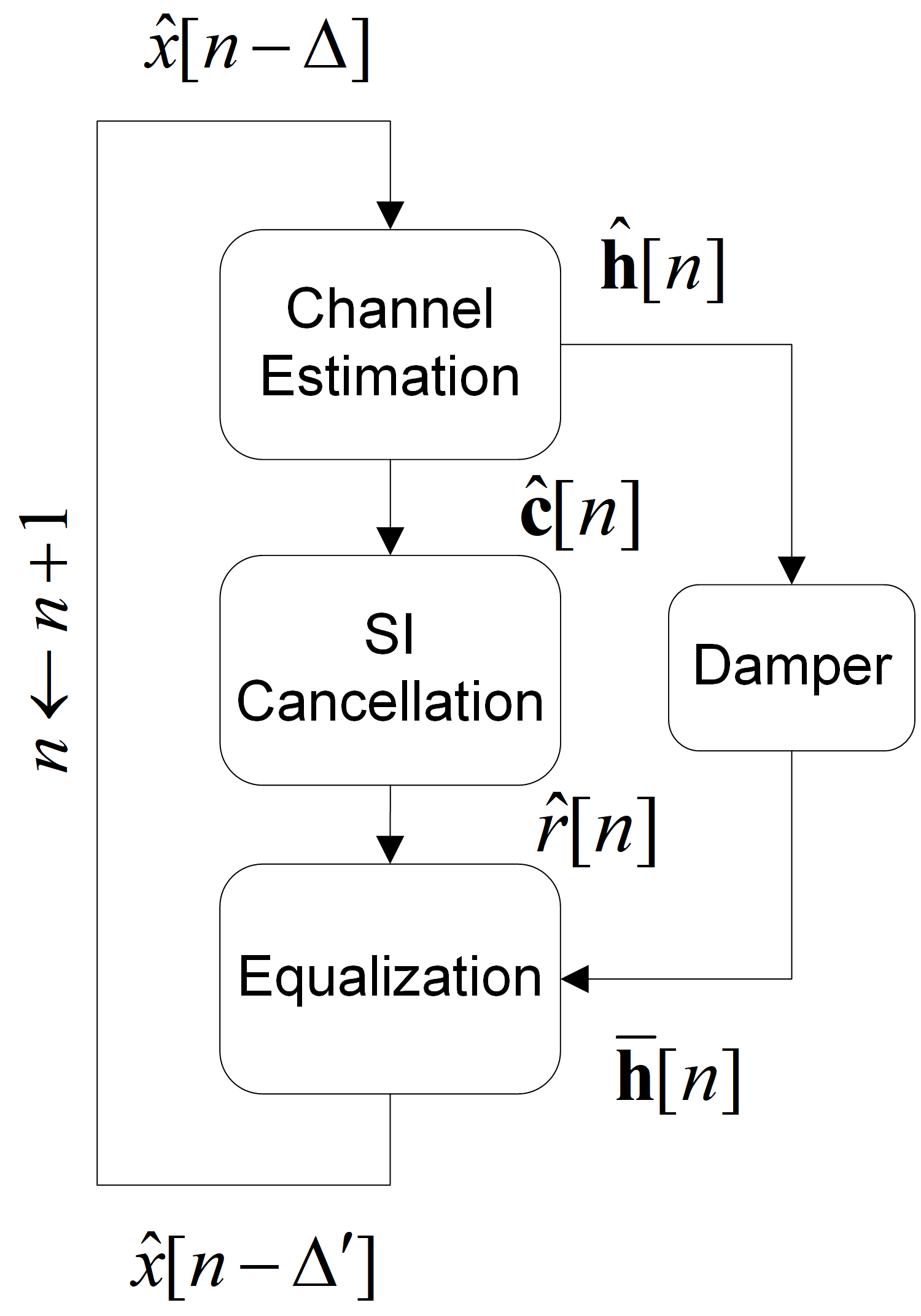}
\caption{Proposed UWA-FD receiver at the $n$th cycle.}
\vspace{-10pt}
\label{F2}
\end{figure}

The block diagram in Fig.~\ref{F2} depicts the proposed UWA-FD receiver.  Specifically, at the $n\text{th}$ cycle, the following tasks are performed:
\textit{i)}	Estimations of both the SI and the remote channel vectors (i.e. $\mathbf{\hat{c}}[n]$ and $\mathbf{\hat{h}}[n]$) are simultaneously updated.
\textit{ii)} Using the provided SI channel estimate from the first stage, the $n$th sample of the SI signal, $\hat{s}[n]$, is estimated  and eliminated from the received signal $y[n]$ to obtain the SI-free signal $\hat{r}[n]$.
\textit{iii)} The estimated remote channel from the first stage, passes through a damper, which is used to suppress the undesired fluctuations of the estimates. 
By using a damped version of the estimated remote channel, the SI-free signal, $\hat{r}[n]$, is equalized to detect the data symbol $\hat x[n - \Delta ']$,  where $\Delta '$ is the delay of the equalizer.  Finally, the detected data symbol is used as a reference in the channel estimation stage at the next cycle with $\Delta=\Delta'+1$ delays. 
In the following, we individually explain the stages of the proposed receiver. 

\vspace{-5pt}
\subsection{Channel Estimation Stage}
Channel estimation is one of the main concerns in UWA-FD systems; with a precise estimate of the channels, the receiver is able to accurately remove the SI and, then, detect the data symbols from the SI-free signal. However, it must be noted that, since the SI channel is significantly stronger than the remote channel, the accuracy of the SI channel estimation is more critical.  In addition, because of the time-variations of the UWA transmission media, the channel estimator must provide a real-time estimate  of both the SI and remote channels in order to keep track of the variations \cite{R10}. 

As mentioned before, the SI and remote channels are jointly estimated at the channel estimation stage. At this stage, the known signal $i[n]$ is used as the reference sequence for estimating the SI channel; while, for estimating the remote channel, the remotely transmitted data symbol, $x[n]$, is the reference. However, since $x[n]$ is unknown, we use the previously detected data symbols as the reference. 

Amongst the various adaptive methods, we use the recursive least squares (RLS) algorithm which has been shown to be a good estimator for time-varying channels \cite{R12}. At the $n\text{th}$ cycle, the input of the RLS algorithm includes the received signal $y[n-\Delta]$, the SI reference sequence $\mathbf{i}[n-\Delta]$, and the previously detected data symbols $\mathbf{\hat{x}}[n-\Delta ]=\left[ \hat{x}[n-\Delta ],\ldots \hat{x}[n-\Delta -L+1] \right]^{T}$.

The reference sequences $\mathbf{\hat{x}}[n-\Delta ]$ and $\mathbf{i}[n-\Delta ]$ are nested in a single column vector as ${\bf{v}}[n - \Delta ] = {[{\bf{\hat x}}[n - \Delta ]^{T}\,\,{\mkern 1mu} {\mkern 1mu} {\mkern 1mu}  \vdots \,\,{\mkern 1mu} {\mkern 1mu} {\mkern 1mu} {\bf{i}}[n - \Delta ]^{T}]^T}$ and the estimates of the channel vectors are also concatenated in a column vector as  ${\bf{u}}[n - \Delta ] = {[{{\bf{\hat h}}^H}[n - \Delta ]\,\,\,\, \vdots \,\,\,{{\bf{\hat c}}^H}[n - \Delta ]]^H}$. The aim of the RLS estimator is to update the previous vector $\mathbf{u}[n-\Delta -1]$ and obtain $\mathbf{u}[n-\Delta ]$. To this end, the estimated received signal is (see \eqref{1} and \eqref{2})
\begin{equation}
\hat{y}[n-\Delta ]={{\mathbf{u}}^{H}}[n-\Delta -1]\,\mathbf{v}[n-\Delta].
\label{3}
\end{equation}
Then, the error $e=y[n-\Delta ]-\hat{y}[n-\Delta ]$ is used to update $\mathbf{u}[n-\Delta -1]$. The RLS algorithm for the joint channel estimation at the $n\text{th}$ cycle is presented in Algorithm 1, in which $\lambda$ is the forgetting factor and $\delta$ is a small scalar for initiation.

\begin{algorithm}[t!]
\caption{: RLS algorithm to jointly estimate the SI and remote channels at the $n$th cycle}
\begin{algorithmic}
\IF {$n=0$}
			\STATE Set $\lambda$ and $\delta$
			
			\STATE Initiate ${\bf{P}} \leftarrow \frac{1}{\delta }{{\bf{I}}_{M + L}}$

\ELSE 

      \STATE Input : ${\bf{u}}[n - \Delta  - 1]$ and ${\bf{v}}[n - \Delta ]$
			
      \STATE $\hat y[n - \Delta ] \leftarrow {{\bf{u}}^H}[n - \Delta  - 1]\,{\bf{v}}[n - \Delta ]$
			
	    \STATE $e  \leftarrow y[n - \Delta ] - \hat y[n - \Delta ]$
			
	    \STATE ${\bf{\gamma }} \leftarrow {(\lambda  + {{\bf{v}}^H}[n - \Delta ]\,{\bf{Pv}}[n - \Delta           ])^{ - 1}}{\bf{Pv}}[n - \Delta ]$
			
	    \STATE ${\bf{P}} \leftarrow {\lambda ^{ - 1}}\left( {{\bf{I}} - {\bf{\gamma }}{{\bf{v}}^H}[n -        \Delta ]} \right)\,{\bf{P}}$

			\STATE  ${\bf{u}}[n - \Delta ] \leftarrow {\bf{u}}[n - \Delta  - 1] + {e^*}{\bf{\gamma }}$
			
	   \STATE Return :  ${\bf{u}}[n - \Delta ]$		
\ENDIF	
\end{algorithmic}
\end{algorithm}

According to \eqref{3}, because of the equalizer delay in data symbol detection, the channel estimation is also delayed, such that, at the $n\text{th}$ cycle, the channel vectors at time $n-\Delta $ are estimated (i.e. $\mathbf{\hat{c}}[n-\Delta ]$ and $\mathbf{\hat{h}}[n-\Delta ]$). In other words, the channel estimation stage is $\Delta $ time instances behind the SI cancellation and equalization stages, where $\mathbf{\hat{c}}[n]$ and $\mathbf{\hat{h}}[n]$ are required, respectively. However, recall from \eqref{1} and \eqref{2} that the time-variations are assumed to be slow enough so that the SI and remote channels remain constant over, at least,  $K=\max \{ M,L\}$ time instances (i.e. $\mathbf{c}[n]\approx \mathbf{c}[n-K]$ and $\mathbf{h}[n]\approx \mathbf{h}[n-K]$).  Therefore, by adjusting $\Delta $ to satisfy $0\le \Delta \le K$, we can use the approximations $\mathbf{\hat{c}}[n]\approx \mathbf{\hat{c}}[n-\Delta ]$ and $\mathbf{\hat{h}}[n]\approx \mathbf{\hat{h}}[n-\Delta ]$, so that the estimated channel at the channel estimation stage can be utilized at the next two stages. 

As mentioned above, in the RLS algorithm, both the SI and remote signals are used to construct $\hat{y}[n-\Delta ]$, as shown in \eqref{3}. This implies that for estimating the SI channel, we not only use the strong SI signal, but we also take the history of the weak remote signal into account. In the simulations, we will show that the proposed procedure results in a more accurate SI channel estimation compared to other adaptive receivers in which the remote signal is treated as an additive noise during the SI channel estimation. 

The other point is that, in the proposed RLS algorithm, there is no need to transmit frequent training sequences from the remote transmitter; instead, we use the previously detected data symbols as the reference for estimating the remote channel. Hence, the bandwidth efficiency is preserved in the proposed receiver. However, to get a faster initial convergence, in practice, we use a short training sequence to initiate the RLS algorithm. Based on the simulation results, the length of the initial training should be a little longer than $M+L$ symbols, which is small and does not significantly reduce the bandwidth efficiency of the proposed receiver during a long-term utilization of the system. 

\vspace{-5pt}
\subsection{Self-Interference Cancellation Stage}
The next stage, as shown in Fig.~\ref{F2}, is SI cancellation. By using the SI reference sequence and the estimate of the SI channel, provided from the first stage, the SI signal is reconstructed and subtracted from the received signal as	
\begin{align}
  & \hat{s}[n]={{{\mathbf{\hat{c}}}}^{H}}[n]\,\mathbf{i}[n]; \nonumber\\ 
 & \hat{r}[n]=y[n]-\hat{s}[n],
\label{4}
\end{align}
where $\hat{s}[n]$ and $\hat{r}[n]$ are the estimates of the SI and the SI-free signals, respectively. Since the power of the SI signal is significantly higher than that of the remote signal, it is clear that an accurate estimate of the SI channel is required in \eqref{4}. 
To provide some intuition, consider an example in which the power of $i[n]$ and $x[n]$ are ${{P}_{i}}={{P}_{x}}=0$ dB. The power of the SI and remote channel vectors are $\text{E}\left\| \mathbf{c}[n] \right\|^{2}=10$ dB and $\text{E}\left\| \mathbf{h}[n] \right\|^{2}=-20$ dB, respectively. Therefore, the power of the SI and remote signals at the receiver become $P_s=10$ dB and $P_r=-20$ dB, respectively.
A small deviation between the actual and estimated SI channel,  $\text{E}\left\| \mathbf{c}[n]-\mathbf{\hat{c}}[n] \right\|^{2}=-20$ dB, results in a residual SI with power ${{P}_{\text{res} }}=-20$ dB, which is in the same range as the remote signal power, $P_r$. This significant residual SI, caused by a small inaccuracy in the estimated channel, will significantly deteriorate the performance. 

\vspace{-5pt}
\subsection{Equalization Stage}
The last stage in the proposed receiver is equalization, in which the estimate of the SI-free signal, $\hat r[n]$, is equalized with the purpose of eliminating the multipath effects of the remote channel from the data symbols. In the proposed receiver, we use a decision feedback equalizer (DFE), which is a non-linear equalizer; by employing a feedforward filter (FFF) and a feedback filter (FBF), a DFE outperforms the linear zero forcing (ZF) and minimum mean squared error (MMSE) equalizers in frequency selective channels \cite{R7}. In a DFE, the FFF with length $\mathcal{L}$ acts as linear equalizer for the first $\mathcal{L}$ paths of the channel and the FBF removes the inter-symbol-interference caused by the rest of the paths and the FFF. Also a DFE with these properties, leads to a $\Delta'  = {\cal L} - 1$ symbol delay. So, at the $n$th cycle, the output of the equalization stage is $\hat{x}[n-\Delta' ]$, which is the detected version of $x[n-\Delta' ]$.

Since the DFE removes the effects of multipath  for remote channel ${\bf{h}}[n]$, the FFF and FBF weights are tuned based on the estimate of this channel \cite{R8}. Recall that the estimation of the remote channel, ${\bf{\hat h}}[n]$, is already provided from the channel estimation stage. However, because of possible errors in the estimated remote channel, the direct use of ${\bf{\hat h}}[n]$ to tune the DFE filters is not optimal. Since the remote channel is much weaker than the SI channel, in the joint channel estimation by the RLS algorithm, its estimate is susceptible to the noise and the errors at the previously detected data symbols. These unwanted effects lead to incorrect and artificial fluctuations in ${\bf{\hat h}}[n]$ at each cycle. Hence, by directly using ${\bf{\hat h}}[n]$ to tune the DFE, the noise and symbol errors propagate to the next cycles and deteriorate the performance.

In order to prevent this damage and suppress the propagation of the artificial fluctuations in ${\bf{\hat h}}[n]$ to the next cycles, we use ${\bf{\bar h}}[n]$ to tune the FFF and FBF at the DFE, where ${\bf{\bar h}}[n]$ is the damped version of the estimated channel and is given as
\begin{equation}
{\bf{\bar h}}[n]=(1-{{\mu }}){\bf{\bar h}}[n-1]+{{\mu }}{\bf{\hat h}}[n],
\label{6}
\end{equation}
where $0\le {{\mu }}\le 1$ is the damping factor, chosen so that the artificial and rapid fluctuations in ${\bf{\hat h}}[n]$ are damped; however, the slower natural changes of the channel are not suppressed because the adjustment of the DFE filters is supposed to follow the natural time-variations of the channel at each cycle.
Having said that, one can see from \eqref{6} that if ${{\mu }}=0$, then ${\bf{\bar h}}[n]={\bf{\bar h}}[n-1]$, which means that any changes, including the artificial and natural changes, at ${\bf{\hat h}}[n]$ are extremely damped and ${\bf{\bar h}}[n]$ remains constant. On the other hand, if ${{\mu }}=1$, ${\bf{\bar h}}[n]={\bf{\hat h}}[n]$, indicating that no damping is imposed.
So, the value of ${{\mu }}$ is determined based on the speed of the natural time-variations of the actual remote channel ${\bf{ h}}[n]$. If ${\bf{h}}[n]$ is changing slowly with a long coherence time, ${{\mu }}$ is set to a small value; however, if ${\bf{h}}[n]$ is rapidly time-varying with a shorter coherence time, ${{\mu }}$ is set to a larger scalar.
 
\vspace{-5pt}
\section{Simulation Results}
In this section we numerically evaluate the performance of the proposed receiver for UWA-FD system. We consider a bandwidth $B=5\,\text{kHz}$ and use normalized power BPSK modulation at both the local and remote ends.

The simulations are performed in baseband; however, in order to model the non-linearity of the local PA,
we first generate the passband input of the PA and then, downconvert the output to baseband. To this end, we consider a carrier frequency ${{f}_{\text{c}}}=12\,\,\text{kHz}$ and a root raised cosine filter, with the roll-off-factor $\alpha =0.5$, to shape the signal; the length of the filter is truncated to $12$ symbols duration. The odd harmonics of the Taylor series expansion are used to relate the input (i.e. $p(t)$) and output (i.e. $q(t)$) of the PA as $q(t)=\sum\limits_{m=1,3,5}{{{a}_{m}}p{{(t)}^{m}}}$ \cite{R6}, where ${{a}_{1}}=100$, ${{a}_{3}}=5$ and ${{a}_{5}}=10$. The local PA noise, ${w_{{\rm{PA}}}}(t)$ has a Gaussian distribution with power $\sigma _{\text{PA}}^{2}=10\,\text{dB}$.
Then, the local transmitter reference sequence, $i[n]$, is the downconverted, matched filtered,  and downsampled version of $i(t)=q(t)+{w_{{\rm{PA}}}}(t)$, such that the power of $i[n]$ is normalized to $P_i=0$ dB.

\begin{figure}[t!]
\centering
\includegraphics [width=2.3in]{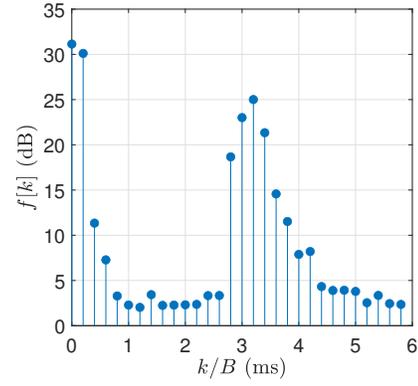}
\caption{The estimated PDP for the SI channel in a lake experiment \cite{R10}. The direct path (first peak) is stable with a long coherence time; however, the reflected path from the surface (second peak) is time-varying with a coherence time of $70$ ms.}
\vspace{-10pt}
\label{F3}
\end{figure}

The baseband power delay profile (PDP) of the SI channel is $\text{PDP}_{{{\rm{SI}}}}[k] = {\beta _1}{\mkern 1mu} f[k]$, where $f[k]$, shown in Fig.~\ref{F3}, is the estimated PDP for the SI channel in a lake experiment in \cite{R10}. The remote channel is assumed to have 
an exponential PDP as $\text{PDP}_{{\text{remote}}}[k]={{\beta }_{2}}\,{{e}^{-0.25\,k}}$. The factors ${{\beta}_{1}}$ and ${{\beta}_{2}}$ are chosen so that the power of the SI and remote signals are ${{P}_{s}}$ and ${{P}_{r}}$, respectively. The lengths of the SI and remote channels are  $M=30$ and  $L=70$, respectively. All paths of both channels are time-varying with a coherence time $70$ ms, except the direct path of the SI channel which is relatively time-invariant \cite{R10}.

At the receiver side, the forgetting factor for the RLS algorithm is set to $\lambda =0.98$, and $\delta =0.0001$. At the DFE, we take the FFF length as ${\cal L}=70$ and the FBF length as 50. With this consideration,  the equalizer delay for delivering the data symbols becomes $\Delta' =69$.  Based on the coherence time of the remote channel, we choose the damping factor as  ${\mu} = {10^{ - 3}}$. In addition, the initial training sequence from the remote transmitter contains $130$ symbols, or $130/B=26\,\text{ms}$.

We assume the powers of the SI signal, remote signal and ambient noise are ${{P}_{s}}=0\,\text{dB}$, ${{P}_{r}}=-20\,\,\text{dB}$ and $\sigma _0^2 =  - 35$ dB, respectively.
We compute the normalized mean square error (MSE) for the SI channel estimation in the proposed receiver (in which the SI and remote channels are jointly estimated) and the conventional receiver (where the SI channel is estimated by ignoring the remote signal) as 
\begin{equation}
\begin{array}{l}
{\rho _{{\bf{\hat c}}}} = \frac{{{\rm{E}}{{\left\| {{\bf{c}}[n] - {\bf{\hat c}}[n]} \right\|}^2}}}{{{\rm{E}}{{\left\| {{\bf{c}}[n]} \right\|}^2}}} = 9.75 \times {10^{ - 5}};\\
\\
{\rho _{{\bf{\tilde c}}}} = \frac{{{\rm{E}}{{\left\| {{\bf{c}}[n] - {\bf{\tilde c}}[n]} \right\|}^2}}}{{{\rm{E}}{{\left\| {{\bf{c}}[n]} \right\|}^2}}} = 2.9 \times {10^{ - 3}},
\end{array}
\label{7}
\end{equation}

\noindent where ${{\bf{\hat c}}}[n]$ and ${{\bf{\tilde c}}}[n]$ are the SI channel vector estimated by the proposed and the conventional receivers, respectively. 
Clearly, ${\rho _{{\bf{\hat c}}}}$ is much smaller than ${\rho _{{\bf{\tilde c}}}}$, demonstrating the advantage of the proposed receiver in tracking the SI channel and, eventually, in canceling the SI . 

\begin{figure}[t!]
\centering
\includegraphics [width=3.5in]{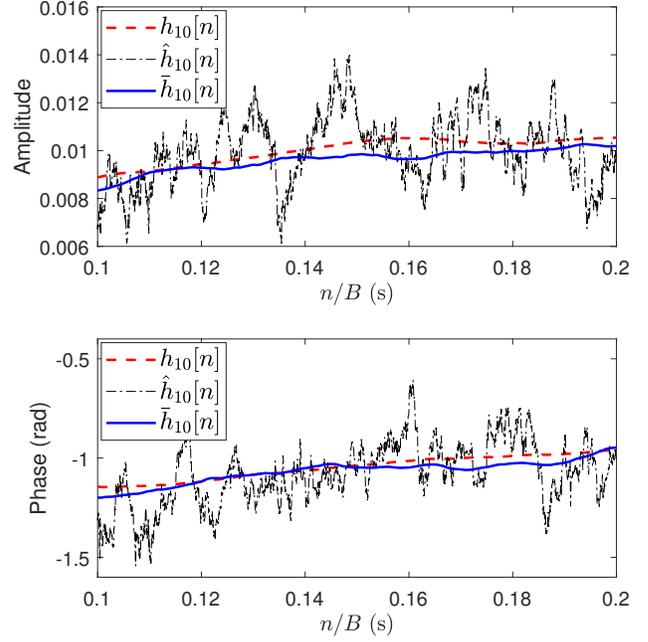}
\caption{Amplitude and phase of the $10$th path of the true (dashed red line), estimated (dash-dotted black line), and damped (solid blue line) remote channels. } 
\vspace{-10pt}
\label{F4}
\end{figure}

In order to illustrate the damper's role in suppressing the artificial fluctuations and reducing the remote channel estimation errors, as an example, we have plotted one realization of the damper's input (which is the estimated remote channel by the RLS algorithm) and output (the channel that is used to tune the DFE) in Fig.~\ref{F4}. This figure shows the $10$th estimated and damped 
delay path of the remote channel denoted by ${{\hat h}_{10}}[n]$ and ${{\bar h}_{10}}[n]$, respectively, and compares them with that of the  true remote channel denoted by $h_{10}[n]$. As seen, the RLS estimated channel has artificial fluctuations over the subsequent cycles. However, damping this estimate kills these artificial fluctuations and passes the smooth and natural time-variations. By looking at Fig.~\ref{F4}, one can intuitively see that the channel estimation error for the damper output is lower than that of the input. To evaluate that, we calculate the  normalized MSE for the input and output of the damper, 
\begin{equation}
\begin{array}{l}
{\rho _{{\bf{\hat h}}}} = \frac{{{\rm{E}}{{\left\| {{\bf{h}}[n] - {\bf{\hat h}}[n]} \right\|}^2}}}{{{\rm{E}}{{\left\| {{\bf{h}}[n]} \right\|}^2}}} = 1.6 \times {10^{ - 2}},\\
\\
{\rho _{{\bf{\bar h}}}} = \frac{{{\rm{E}}{{\left\| {{\bf{h}}[n] - {\bf{\bar h}}[n]} \right\|}^2}}}{{{\rm{E}}{{\left\| {{\bf{h}}[n]} \right\|}^2}}} = 1.6 \times {10^{ - 3}}.
\end{array}
\label{8}
\end{equation}

\noindent Comparing ${\rho _{{\bf{\bar h}}}}$ with  ${\rho _{{\bf{\hat h}}}}$ indicates that the remote channel estimation with RLS suffers from a significant error.  As discussed in Section III-C, this large error is caused by the noise and errors in the previously detected data symbols. However, the relatively lower error in the output of the damper confirms that using a damper corrects the remote channel estimation to some extent, before it is used to tune the DFE.

\begin{figure}[t!]
\centering
\includegraphics [width=3in]{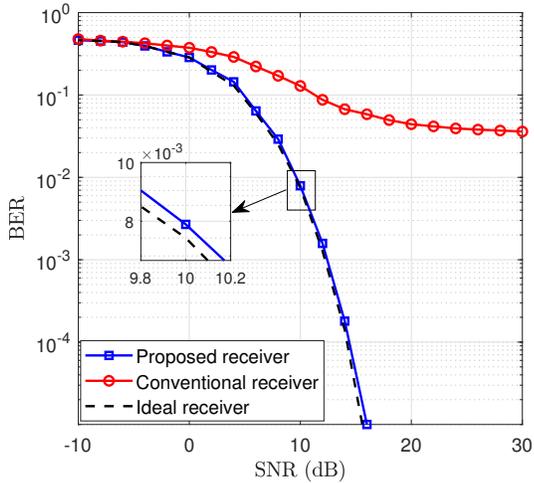}
\caption{BER comparison for the proposed, conventional and ideal receivers ($P_s=0$ dB and $P_r=-20$ dB).} 
\vspace{-5pt}
\label{F5}
\end{figure}

In Fig.~\ref{F5}, we present the BER versus the remote SNR (i.e. $P_r/\sigma _0^2$),  when $P_s=0$ dB, $P_r=-20$ dB and $\sigma _0^2$ is in the range from $-50$ to $-10$ dB. In this figure, the BER of the proposed and conventional receivers are compared with that of the ideal receiver, in which both the SI and remote channel state information is assumed to be perfectly known. Since in the ideal  receiver the SI channel is perfectly known, the SI is completely removed and the SI-free signal is equalized based on the perfect knowledge of the remote channel. 
By looking at Fig.~\ref{F5}, one can see that the BER of the proposed receiver is almost identical to that of the ideal receiver, implying that the proposed receiver can track the variations of both SI and remote channels very well. However, the performance of the conventional receiver is significantly deteriorated. 
This is because of two reasons. First, in this receiver, the SI channel estimation is not as accurate as in the proposed receiver (as shown in \eqref{7}) and a considerable residual SI contaminates the output of the SI cancellation stage.  Second, the variation of the remote channel is not tracked so that the estimate for the remote channel remains fixed to that obtained by using the initial training sequence.

In Figs. 6 and 7, we show the BER when $P_s= 5$ and $10$ dB, respectively. As seen, the BER of the proposed receiver degrades as the SI power increases. 
As a matter of fact, for an even stronger SI, the SI cancellation residual is larger and leads to performance degradation. In order to investigate the impact of the SI power on the residual, in Fig.~\ref{F8}, we plot ${\rho _{\hat r}}$, the normalized MSE of $\hat r[n]$, versus the SI to remote signal power ratio ($P_s/P_r$). 
We assume $P_r=-20$ dB and $\sigma _0^2=-35$ dB, and the $P_s$ range is $-20$ to 20 dB. The normalized MSE for $\hat r[n]$ is calculated as
\begin{equation}
{\rho _{\hat r}} = \frac{{{\rm{E|}}r[n] - \hat r[n]{|^2}}}{{{\rm{E|}}r[n]{|^2}}}.
\label{9}
\end{equation}
This quantity measures the normalized residual power after SI cancellation. 
As seen from Fig.~\ref{F8}, for $P_s/P_r \le 20$ dB, ${\rho _{\hat r}}$ is very low for the proposed receiver, indicating that the performance of this receiver is identical to that of the ideal receiver (as seen in Fig.~\ref{F5}, where $P_s/P_r=20$ dB); however, for $P_s/P_r>20$ dB, a stronger residual results, which leads to a higher BER (as seen in Figs. 6 and 7, where $P_s/P_r =25$ dB and $30$ dB, respectively). 
In addition, ${\rho _{\hat r}}$ of the conventional receiver for $P_s/P_r \le 20$ dB is a significant constant value and surges for $P_s/P_r>20$ dB. For $P_s/P_r=40$ dB, ${\rho _{\hat r}}$ for the proposed receiver equals to that of the conventional receiver.
\begin{figure}[t!]
\centering
\includegraphics [width=3in]{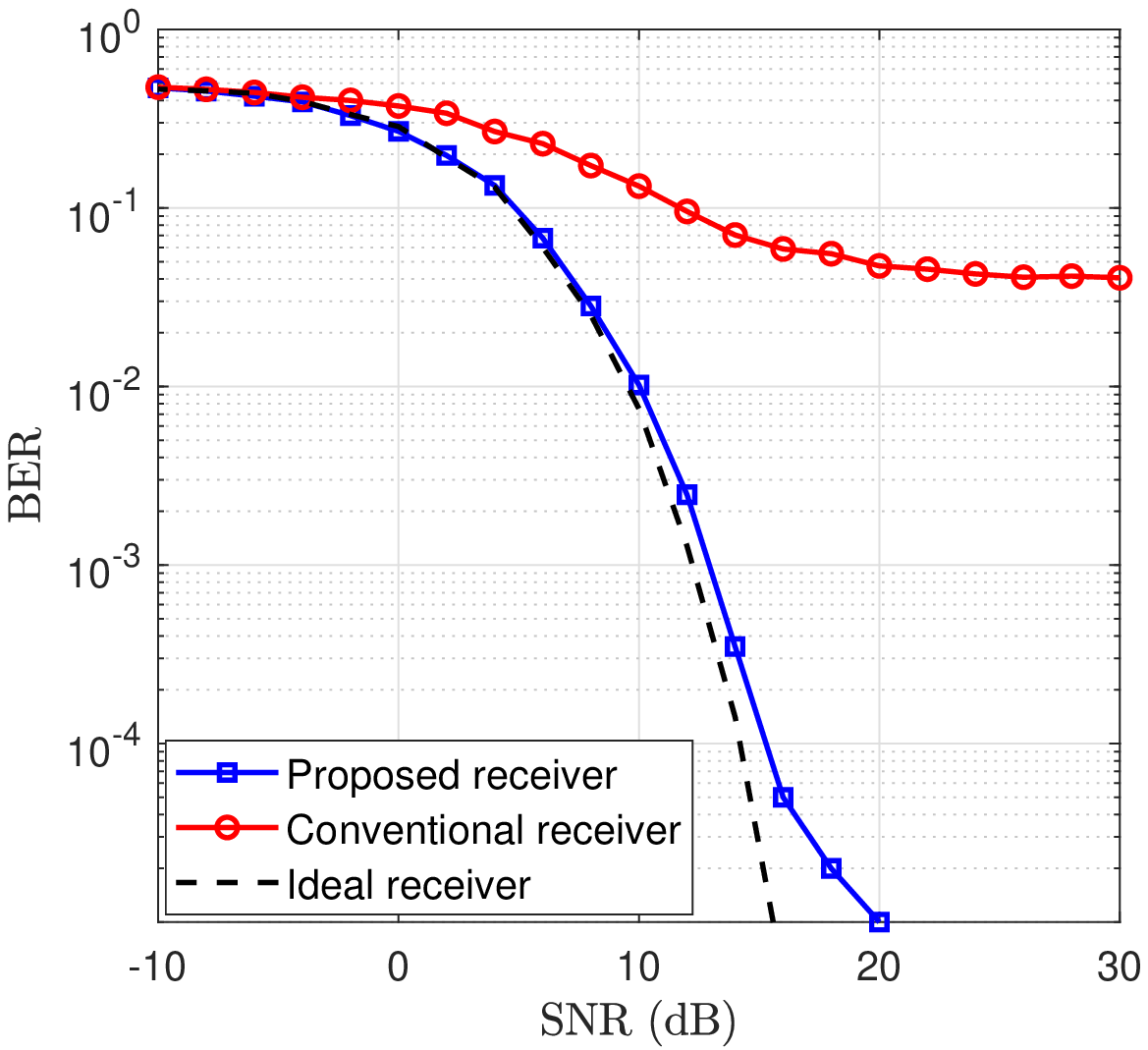}
\caption{BER comparison for the proposed, conventional and ideal receivers ($P_s=5$ dB and $P_r=-20$ dB).}
\vspace{-5pt}
\label{F6}
\end{figure}
\begin{figure}[t!]
\centering
\includegraphics [width=3in]{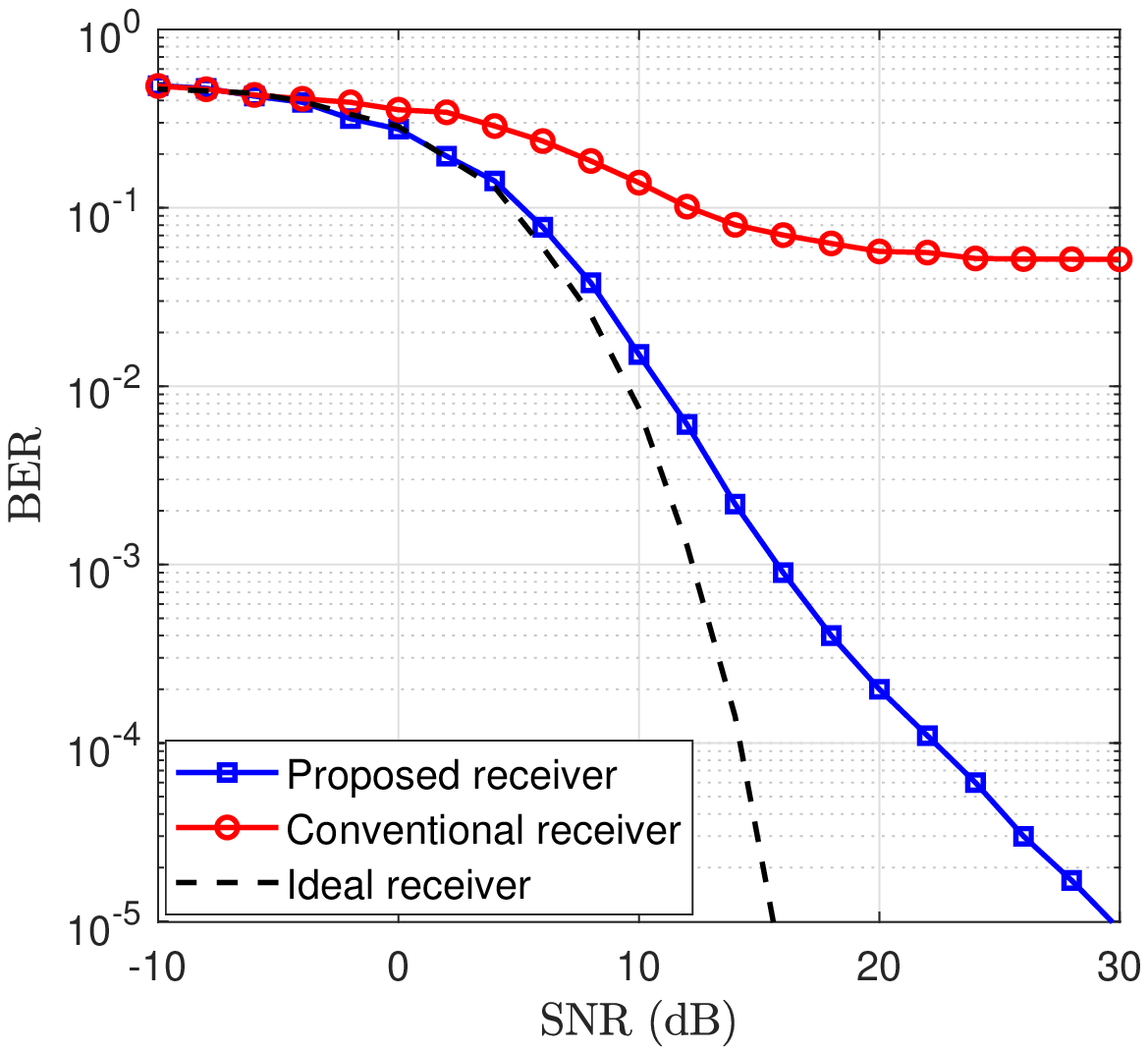}
\caption{BER comparison for the proposed, conventional and ideal receivers ($P_s=10$ dB and $P_r=-20$ dB).}
\vspace{-5pt}
\label{F7}
\end{figure}
\begin{figure}[t!]
\centering
\includegraphics [width=3in]{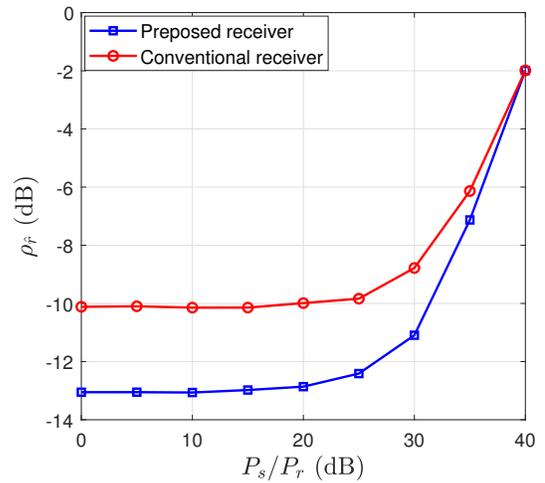}
\caption{Comparison on ${\rho _{\hat r}}$ versus $P_s/P_r$ for the proposed and conventional receivers ($P_r=-20$ dB and $\sigma _0^2=-35$ dB).} 
\vspace{-10pt}
\label{F8}
\end{figure}

\vspace{-5pt}
\section{Conclusions}
In this paper, we proposed a novel receiver for UWA-FD communication with joint tracking of the SI and remote channels. In order to remove the need to frequently send training sequences, the channel estimator is fed by feedback from the previously detected data symbols as a reference for the remote channel. With this strategy, an accurate SI channel estimation is obtained; also, the  remote channel variations are tracked along with the SI channel without using extra training.  
The concern about error propagation due to using feedback is addressed by employing a damper to suppress the harmful effects of the incorrectly detected data symbols in the estimated remote channel. Then, the output of the channel damper is used to tune the equalizer. Simulation results show that the proposed receiver outperforms conventional methods which only track the SI channel. The performance of the proposed receiver decreases as the SI power increases beyond a specific value. This value is determined based on the channel PDPs, the coherence time and the noise level.

\bibliographystyle{IEEEtranTCOM}
\vspace{-15pt}
\bibliography{bare_jrnl_V2}

\end{document}